\documentclass[twocolumn]{aastex701}
\usepackage[utf8]{inputenc}
\usepackage{amsmath}
\usepackage{ulem}
\usepackage{hyperref}
\usepackage{CJK}
\begin{document}
\begin{CJK*}{UTF8}{gbsn}
\title{H\,{\sc i} Content of Group Galaxies from the FAST All Sky H\,{\sc i} Survey}
\correspondingauthor{Qingzheng Yu}
\email{qingzheng.yu@unifi.it}
\correspondingauthor{Taotao Fang}
\email{fangt@xmu.edu.cn}
\author[orcid=0000-0002-3940-2950]{Shulan Yan (鄢淑澜)}
\email{yansl@stu.xmu.edu.cn}
\affiliation{Department of Astronomy, Xiamen University, Xiamen, Fujian 361005, People's Republic of China}
\author{Andrew Ma}
\email{andy.ma@student.isb.bj.edu.cn}
\affiliation{International School of Beijing, 10 An Hua Street, Shunyi District, Beijing 101318, People's Republic of China}
\author[orcid=0000-0003-3230-3981]{Qingzheng Yu (余清正)}
\email{qingzheng.yu@unifi.it}
\affiliation{Dipartimento di Fisica e Astronomia, Universit\`a degli Studi di Firenze, Via G. Sansone 1, 50019 Sesto Fiorentino, Firenze, Italy}
\affiliation{INAF - Osservatorio Astrofisico di Arcetri, Largo E. Fermi 5, I-50125 Firenze, Italy}
\author[orcid=0000-0002-2853-3808]{Taotao Fang (方陶陶)}
\email{fangt@xmu.edu.cn}
\affiliation{Department of Astronomy, Xiamen University, Xiamen, Fujian 361005, People's Republic of China}
\author[orcid=0000-0003-1761-5442]{Chuan He (何川)}
\email{hechuan@xmu.edu.cn}
\affiliation{Department of Astronomy, Xiamen University, Xiamen, Fujian 361005, People's Republic of China}
\author[orcid=0000-0001-6083-956X]{Ming Zhu (朱明)}
\email{mz@nao.cas.cn}
\affiliation{National Astronomical Observatories, Chinese Academy of Sciences (NAOC), Beijing 100101, People's Republic of China}
\affiliation{Guizhou Radio Astronomical Observatory, Guiyang 550025, People's Republic of China}
\affiliation{Guizhou Provincial Key Laboratory of Medium and Low Frequency Radio Astronomy Technology and Application, Guiyang 550025, People's Republic of China}

\begin{abstract}
We investigate the atomic gas (H\,{\sc i}) content of galaxies in groups using early data from the FAST All Sky H\,{\sc i} survey (FASHI). Taking advantage of FAST's blind, wide-area coverage and uniform sensitivity, we assemble a sample of $230$ group galaxies belonging to $182$ groups at $z\leq0.03$. These groups were identified using a halo-based group finder, and they have an median membership of $4$ galaxies. We also derived a matched control sample of isolated systems, and apply censored-data modeling to include both detections and non-detections. At fixed stellar mass and color, we find that the global median H\,{\sc i} fraction of group galaxies differs from that of controls by only $-0.04$ dex ($95\%$ CI [$-0.18,\ 0.16$]), indicating at most a mild average offset. The signal is not uniform across populations: satellites are H\,{\sc i}-poor (median $\Delta f_{\mathrm{HI}}=-0.12$ dex), whereas centrals are not H\,{\sc i}-deficient (median $\Delta f_{\mathrm{HI}}=0.13$ dex). Group galaxies located within $0.5R_{180}$ and in denser systems (richness $>10$ or local density $\Sigma>10\ \mathrm{gal\ Mpc^{-2}}$) show stronger negative offsets, whereas galaxies in the outskirts are statistically indistinguishable from the controls. These results refine earlier reports of global group H\,{\sc i} deficiency: with deeper blind data and uniform treatment of upper limits, we show that H\,{\sc i} depletion is primarily confined to satellites and compact cores rather than being ubiquitous across groups.
\end{abstract}

\keywords{\uat{Galaxy evolution}{594} --- \uat{Galaxy groups}{597} --- \uat{Interstellar atomic gas}{833}}

\section{Introduction}

Galaxy groups and clusters serve as key laboratories for studying how environmental conditions shape galaxy evolution. The ram pressure stripping can efficiently remove gas from galaxy outskirts, particularly for infalling galaxies or those interacting with the cosmic web \citep{Boselli06,Chung09,Jaffe15,Stark16}. Long timescale mechanisms like starvation or tidal interactions \citep[e.g.,][]{Larson80,Moore96} can also deplete the cold gas reservoirs of galaxies and suppress the star formation in groups and clusters \citep{Gunn72,Cayatte90,Vollmer01,Chung09,Odekon16,Ai18,Cortese21,Brown21,Loni21,Molnar22,Brown23,Moretti23,Serra23,Luber25,Sorgho25}. Atomic hydrogen (H\,{\sc i}), a major component of the interstellar medium, often extends well beyond the stellar disk, making it a sensitive tracer of environmental processes and the neutral gas content in galaxies \citep{Cayatte94,Hibbard96,Serra12,Romeo20}. Galaxies are usually considered H\,{\sc i} deficient when their H\,{\sc i} content is lower than expected for their size or stellar mass \citep{Jones23,Sorgho25}. Significant H\,{\sc i} deficiencies have been reported in galaxy groups and clusters \citep[e.g.,][]{Haynes84,Huchtmeier97,Verdes01,Williams02,Cortese11,Yoon17,For21,Deb23,Jones23}. \\

Recent H\,{\sc i} observations of $44$ Hickson compact groups (HCGs) with very large array (VLA) and MeerKAT have revealed widespread H\,{\sc i} deficiencies, particularly in systems exhibiting extended H\,{\sc i} structures or in H\,{\sc i} non-detections \citep{Jones23,Sorgho25}. \citet{For21} also reported that nearly half ($\sim46.5\%$) of the galaxies in the Eridanus group are H\,{\sc i} deficient. H\,{\sc i} deficiency is more pronounced in galaxies residing in denser environments or located closer to cluster centers \citep{Solanes01,Odekon16,Hu21,Deb23}. In contrast, some processes can replenish or sustain gas reservoirs. For example, gas-rich minor mergers or accretion from the cosmic web may increase the gas content of low mass group centrals and thereby enhance star formation activity \citep{Janowiecki17}.\\

To investigate diffuse or extended H\,{\sc i} using large sample, several studies have employed blind single-dish H\,{\sc i} surveys to establish scaling relations and to investigate the gas content of group galaxies \citep{Denes14,Catinella18,Zu20,For21,Li22}. H\,{\sc i} scaling relations vary in different works, depending on sample properties, such as optical magnitude, color, stellar mass, halo mass, or morphology \citep{Catinella10,Jones18,Zu20,For21,Lu24}. These differences in scaling relations can result in uncertainties or discrepancies in measured H\,{\sc i} deficiencies in studies of groups and clusters depending on which relation is adopted. \\

Several large H\,{\sc i} surveys have detected a numerous extragalactic H\,{\sc i} sources in the local universe. The H\,{\sc i} Parkes All Sky Survey (HIPASS) covers $71\%$ of the sky in $-90^{\circ}<\mathrm{decl.}<+25^{\circ}30^{'}$, identifying $5374$ H\,{\sc i} sources at $z<0.043$ \citep{Barnes01,Wong06}. Its beam size is $14'.3$ and spectral resolution is $18\ \mathrm{km\ s^{-1}}$. In the northern sky, the Arecibo Legacy Fast ALFA (Arecibo L-band Survey Array) survey (ALFALFA) \citep{Giovanelli05,Haynes11} covers approximately $7000\ \mathrm{deg^{2}}$ in $0^{\circ}<\mathrm{decl.}<+36^{\circ}$ out to $z<0.06$. The beam size is $3'.8\times3'.3$ and the spectral resolution is $10\ \mathrm{km\ s^{-1}}$ at $1420\ \mathrm{MHz}$. It contains $31502$ extragalactic H\,{\sc i} sources. Additionally, the Widefield ASKAP L-band Legacy All-sky Blind Survey (WALLABY) aims to cover $75\%$ of the sky ($-90^{\circ}<\mathrm{decl.}<+30^{\circ}$) to a redshift $z\sim0.26$ \citep{Koribalski20}. Its angular resolution is $30''$ and the spectral resolution is $4\ \mathrm{km\ s^{-1}}$. The sensitivity of WALLABY is $1.6\ \mathrm{mJy\ beam^{-1}}$. But the interferometry has the missing flux issue, thereby losing some mass of the faint gas. \\

The Five-hundred-meter Aperture Spherical radio Telescope (FAST) is a powerful single-dish telescope in the northern sky. The FAST all sky H\,{\sc i} survey (FASHI) project is expected to detect over $100,000$ extragalactic H\,{\sc i} sources covering the sky in $-14^{\circ}<\mathrm{decl.}<+66^{\circ}$ up to $z\sim0.35$ \citep{Zhang24}. FASHI enables the study of extragalactic H\,{\sc i} across previously unexplored regions of the sky. FASHI has a frequency range of $1.0$-$1.5$ GHz, achieving a detection sensitivity of $\sim0.76\ \mathrm{mJy\ beam^{-1}}$ at a velocity resolution of $6.4\ \mathrm{km\ s^{-1}}$ \citep{Jiang19,Jiang20}. The beam size of FASHI is $\sim2^{\prime}.9$. The $3\sigma$ column density of FASHI is $\sim2.25\times10^{18}(1+z)^2\ \mathrm{cm^{-2}}$ across $5$ channels ($32\ \mathrm{km\ s^{-1}}$). The H\,{\sc i} mass limit at $z\sim0$ is $10^6M_\odot$. This performance represents a significant improvement over HIPASS, ALFALFA and WALLABY \citep{Barnes01,Haynes18,Koribalski20}. The early catalog of FASHI has covered $3$ Abell clusters. The sensitivity of FASHI makes it possible to detect faint and diffuse H\,{\sc i} gas that is often missed by early interferometric observations \citep{Zhu21,Liu23,Xu23,Yu23,Zhou23}, offering a clear advantage for probing stripped H\,{\sc i} in group environments. \\

To better understand H\,{\sc i} deficiency in galaxy groups, we cross-match the FASHI data with SDSS DR$7$ group catalog derived from \citet{Yang07} \footnote{\url{https://gax.sjtu.edu.cn/data/Group.html}}. The galaxy groups are selected using a halo-based group finder from \citet{Yang05}. They assign group members based on the properties of the dark matter halo rather than through a traditional friend-of-friend (FOF) algorithm \citep{Davis85}. We selected galaxies based on their optical properties to avoid selection biases toward H\,{\sc i} detections. Most previous studies have focused only on H\,{\sc i} detections or treated non-detections simplistically, potentially introducing bias. A comprehensive assessment requires incorporating upper limits and applying censored-data modeling techniques. Analyses based on H\,{\sc i}-detected samples tend to emphasize gas-rich galaxies and may not reveal the intrinsic H\,{\sc i} deficiency. We also consider galaxies without H\,{\sc i} detections to establish a robust scaling relation, thereby enabling a statistically robust investigation of atomic gas in galaxy groups.  \\

This paper is structured as follows. In Section $2$, we describe the sample selection of galaxies in groups and isolated galaxies as the control sample. Section $3$ presents our main results on H\,{\sc i} gas in group galaxies. H\,{\sc i} deficiency is modest in the full sample. In Section $4$, we expanded the discussion to include the H\,{\sc i} deficiency of both central and satellite galaxies, and we added a comparison with previous studies. Section $5$ summarizes the main results. Throughout this paper, we adopt the standard $\Lambda$-CDM cosmology with $\Omega_{m}=0.3$, $\Omega_{\Lambda}=0.7$, $H_0=70\ \mathrm{km\ s^{-1}Mpc^{-1}}$ and $h=H_0/(100\ \mathrm{km\ s^{-1}\ Mpc^{-1}})$. \\

\section{Method}
\begin{figure*}
    \centering
    \includegraphics[width=0.8\textwidth]{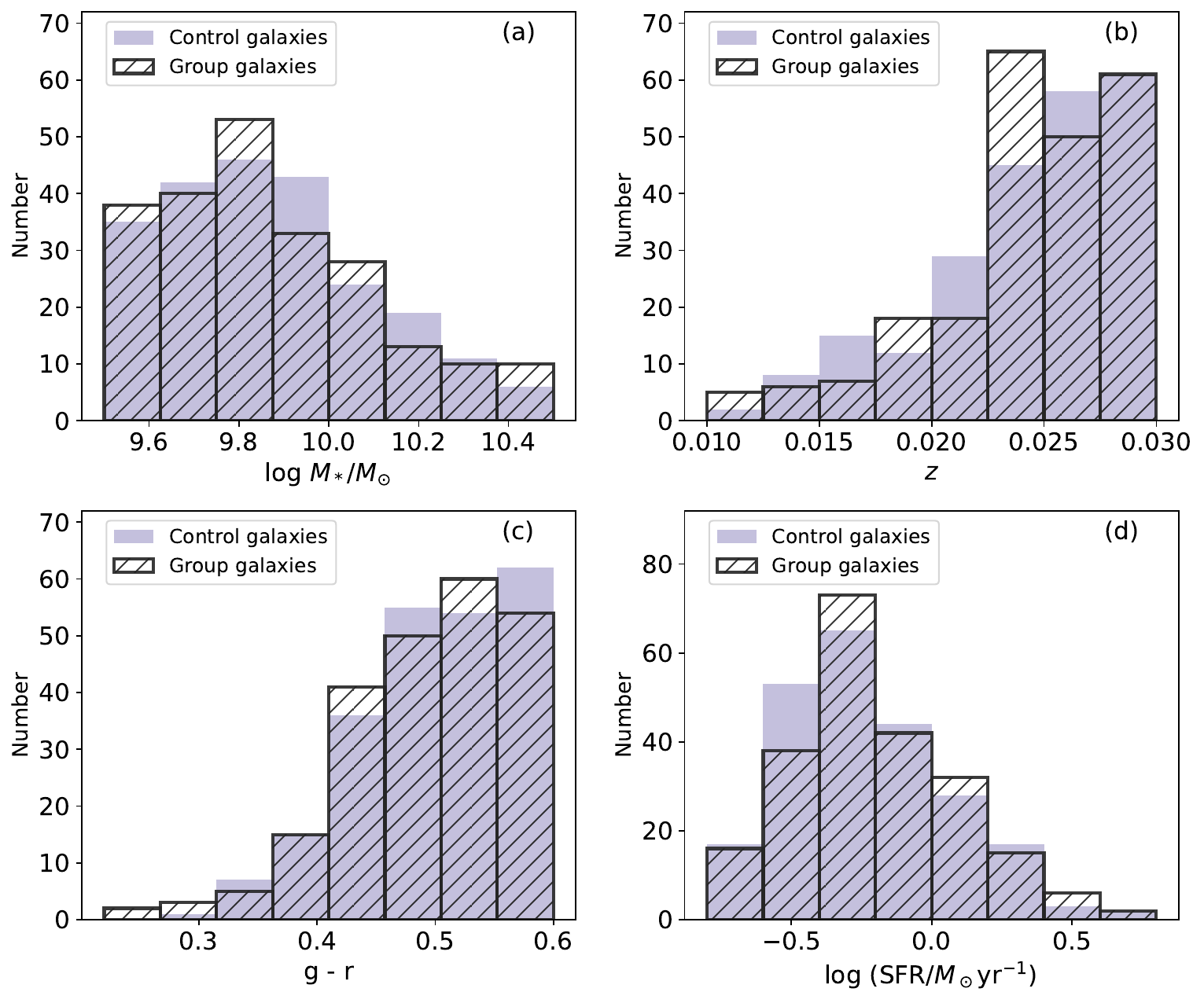}
    \caption{Distributions of basic properties for isolated galaxies (purple) and group galaxies (black). Panel (a), (b), (c) and (d) correspond to stellar mass, redshift, color $g$-$r$ and the SFR, respectively.}
    \label{fig:prop_dist}
\end{figure*}
\subsection{Sample Selection}
Galaxies with $\log(M_*/M_{\odot})\ge9.5$ typically have $M_{\mathrm{HI}} >10^{9.5}M_{\odot}$ \citep{Bradford15}, and the H\,{\sc i} gas fraction decreases with increasing stellar mass \citep{Catinella18}. To avoid massive and gas-poor galaxies that would reduce the detection rate, we restrict the sample to $9.5\le\log(M_*/M_{\odot})\le10.7$ and $r$-band apparent magnitude $m_r<16$ mag. Because red and quiescent galaxies contain little H\,{\sc i} gas \citep{Zuo18,Zu20}, we further limited our sample to blue galaxies with optical colors $g-r\le0.6$ and $\log(\mathrm{SFR})>-1\ M_{\odot}\mathrm{yr^{-1}}$. Stellar mass and star formation rate (SFR) are taken from the Galaxy Evolution Explorer ($GALEX$), Sloan Digital Sky Survey (SDSS), and Wide-field Infrared Survey Explorer (WISE), or $GALEX$-SDSS-$WISE$ Legacy Catalog \citep[GSWLC-X2;][]{Salim16,Salim18}, derived through UV-optical spectral energy distribution fitting. To ensure a high H\,{\sc i} detection rate ($>50\%$) in FASHI, we applied a redshift cut of $z \leq 0.03$. The SDSS spectroscopic completeness is $\sim90\%$ at $z\leq0.03$. The black columns in Figure \ref{fig:prop_dist} show the basic properties of group galaxies, including the stellar mass (a), redshift (b), $g-r$ color (c), and the SFR (d).\\

\subsection{Optical and H\textsc{i} Data}
We adopt the group catalog from \citet{Yang07}. They identify galaxy groups using halo-based group finder from \citet{Yang05}. Unlike the FOF algorithm, this method selects groups associated with the dark matter haloes. This group finder works through the following steps: (1) potential group centers are first identified using the FOF algorithm; (2) the characteristic luminosity is determined by combining the luminosities of all group members; (3) the mass, size, and velocity dispersion of each tentative group are then estimated; (4) group memberships are updated based on these halo properties; and (5) these procedure is iterated until no new members are found. This catalog includes $472,416$ groups within the redshift range $0.01\le z\le0.2$, of which $68,170$ have at least two member galaxies ($44,470$ are pairs), including brightness central galaxies (BCGs). Although the brightest galaxy in loose groups (members $\le3$) may not resemble a real BCG in clusters or compact groups, it still dominates the potential well and thus can reasonably be regarded as the central galaxy. We then identify the brightest galaxy in each group as the central, while the others are classified as satellites. We estimated the local density ($\Sigma$) for our sample by selecting galaxies within $\Delta v \le 1000\ \mathrm{km\ s^{-1}}$. For each target galaxy, we identified the five nearest neighbors and measured the projected distance to the fifth neighbor ($d_5$) \citep{Muldrew12,Schaefer17}. The local density was calculated as $5/(\pi d_5^2)$. The median local density is $0.58\ \mathrm{gal\ Mpc^{-2}}$ for control galaxies and $2.79\ \mathrm{gal\ Mpc^{-2}}$ for group galaxies, respectively. \\

We use the extragalactic H\,{\sc i} data from the FASHI project \citep{Zhang24}. The first data release of the FASHI project covers the sky regions $0^{\mathrm{h}}\leq\mathrm{RA}\leq17.3^{\mathrm{h}}$, $22^{\mathrm{h}}\leq\mathrm{RA}\leq24^{\mathrm{h}}$ and $-6^{\circ}\leq\mathrm{DEC}\leq0^{\circ}$, $30^{\circ}\leq\mathrm{DEC}\leq66^{\circ}$. They found sources using the H\,{\sc i} Source Finding Application (SoFiA) \citep{Serra15,Westmeier21,Westmeier22}. The SoFiA smooths the data over multiple user-defined spatial and spectral scales and measures the noise level at each smoothing iteration. The detection threshold is $4.5\sigma$ in the SoFiA setup. In total, FASHI has detected $41,741$ extragalactic H\,{\sc i} sources at $z<0.09$. More than $94\%$ have a signal-to-noise ratio (SNR) above $10$. The H\,{\sc i} mass detected by FASHI is from $\sim10^{6}M_\odot$ to $\sim10^{11.2}M_\odot$, and the mean value of H\,{\sc i} mass in FASHI is $\sim10^{9.4}M_\odot$.\\

\begin{figure}[h]
    \centering
    \includegraphics[width=0.44\textwidth]{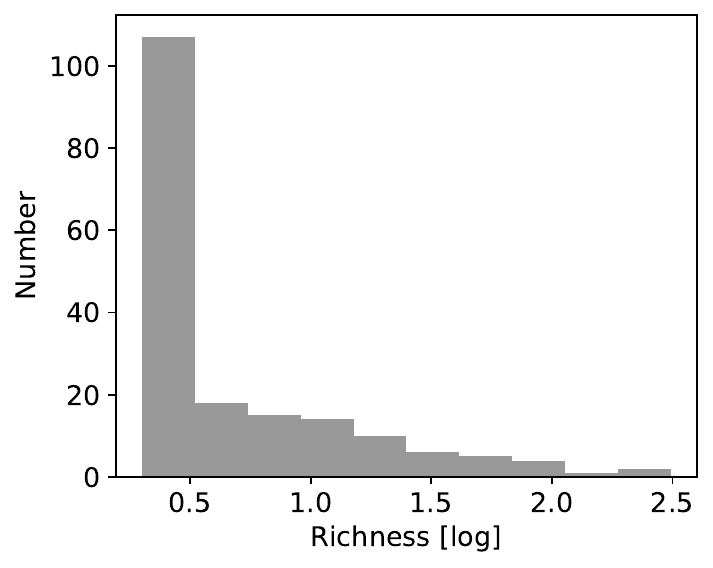}
    \caption{The number distribution of group richness for our selected groups.}
    \label{fig:group_rich}
\end{figure}

To minimize single-beam confusion, we required the nearest optical neighbor to be at least $\sim2^{\prime}.9$ from the target galaxy, so that only one
plausible optical counterpart lies within a FAST beam. This angular separation corresponds to $\sim100\ \mathrm{kpc}$ at $z\sim0.03$, implying that our selection may bias the sample toward small or compact groups at low redshift and large or loose groups at higher redshift. The $2'.9$ separation criterion was adopted solely to avoid single-beam confusion and to ensure unique optical counterparts, rather than to impose any physical isolation threshold. Although this angular cut corresponds to larger projected separations at higher redshift, it is applied uniformly to both the group and control samples and therefore does not bias their relative comparison. We then associated FASHI sources to SDSS targets within $\sim2^{\prime}.9$ and $|\Delta v|\leq 1000\ \mathrm{km\ s^{-1}}$. The final sample contains $230$ group galaxies (belonging to $182$ groups), including $154$ detected in H\,{\sc i} (detection rate $\sim67\%$). Among these, $69$ are classified as centrals ($55$ detected), while $161$ are satellites ($99$ detected). Figure \ref{fig:group_rich} indicates the distribution of group richness and the median number of group members is $4$ in our sample. \\

For galaxies without H\,{\sc i} detection in FASHI catalog, we estimate upper limits following the rms map in FASHI data. The available rms map in FASHI is derived from H\,{\sc i}-detected sources. For each galaxy we selected the closest FASHI source with a velocity difference $|\Delta v|\le1000\ \mathrm{km\ s^{-1}}$. We use the rms of this source and estimate the upper limit following the formula from \citet{Zhang24}:
\begin{equation}
    S_{\mathrm{HI}}=\mathrm{SNR}\times\frac{\sigma_{\mathrm{rms}}}{\sqrt{w_{\mathrm{smo}}}}\times W_{50}\ \mathrm{Jy\ km\ s^{-1}}\ ,
\label{eq:h1Sbf}
\end{equation}
\\
where $w_{\mathrm{smo}}=W_{50}/6.4$ is a smoothing width expressed as the number of spectral resolution bins of $6.4\ \mathrm{km\ s^{-1}}$ bridging half of the signal width. For $L_*$ galaxies, twice the typecal disk rotation velocity is $\sim200\ \mathrm{km\ s^{-1}}$ \citep{Lelli19}. We then adopt $\mathrm{SNR}=5$ and $W_{50}=200\ \mathrm{km\ s^{-1}}$. Finally, we calculated H\,{\sc i} mass via the formula \citep{Meyer04,Ellison18}:
\begin{equation}
\frac{M_{\mathrm{HI}}}{M_\odot}=\frac{2.356\times10^{5}}{1+z}(\frac{D}{\mathrm{Mpc}})^{2}\frac{S_{\mathrm{HI}}}{\mathrm{Jy\ km\ s^{-1}}},
\label{eq:h1mass}
\end{equation}
where $D$ is the distance of source and $z$ is source redshift. H\,{\sc i} gas fraction is calculated using:
\begin{equation}
    f_{\mathrm{HI}}=\frac{M_{\mathrm{HI}}}{M_*}\ ,
\label{eq:fh1}
\end{equation}
where $M_*$ is the stellar mass of the galaxy.\\

\begin{figure}[ht]
    \centering
    \includegraphics[width=0.48\textwidth]{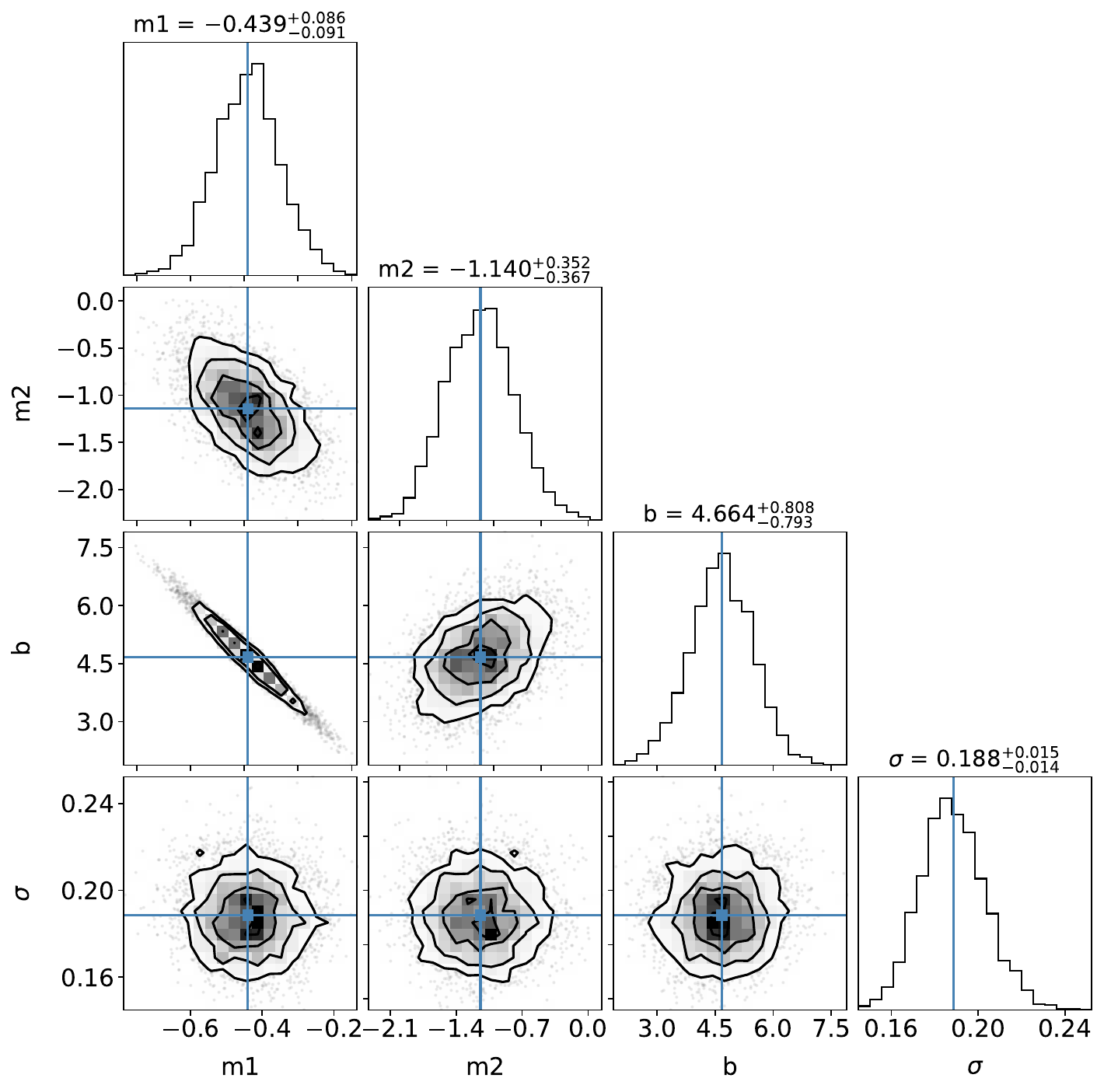}
    \caption{The posterior distributions for the H\,{\sc i} gas fraction prediction: $\log f_{\mathrm{HI}}=m_1\log M_*+m_2(g-r)+b+\sigma_{\mathrm{HI}}$. The blue lines show the mean values for the parameters. The contours in the off-diagonal panels indicate the $68$ percent, $95$ percent, and $98$ percent confidence levels, respectively.}
    \label{fig:mcmc_distri}
\end{figure}

\subsection{HI Relation of Control Sample}

We selected galaxies with group richness $=1$ as the parent isolated galaxy sample and applied the same optical selection criteria used for group galaxies. For galaxies with halo mass below $4\times10^{11}M_\odot$, the halo radius is less than $200\ h^{-1}\mathrm{kpc}$ \citep{Yang07}. Even though these galaxies are identified with group richness $=1$, they might interact with their neighbors located at $200\ h^{-1}\mathrm{kpc}$. Interactions can significantly affect gas content of the galaxies \citep{Yu22,Huang24}. To avoid potential interactions between isolated galaxies, we excluded interacting galaxies following \citet{Feng19}. Interacting galaxies are defined by the line of sight velocity difference of $|\Delta v|\le500\ \mathrm{km\ s^{-1}}$ and projected separation $d_{\mathrm{p}}\le200\ h^{-1}\mathrm{kpc}$. We selected $230$ control galaxies following the distribution of group galaxies, of which $140$ have H\,{\sc i} detections, indicating a detection rate $\sim61\%$. We used Equation \ref{eq:h1Sbf} to \ref{eq:fh1} to calculate H\,{\sc i} mass and gas fraction for non-detections. The purple columns in Figure \ref{fig:prop_dist} indicate the basic properties of the control galaxies. \\

\begin{figure}[h]
    \centering
    \includegraphics[width=0.44\textwidth]{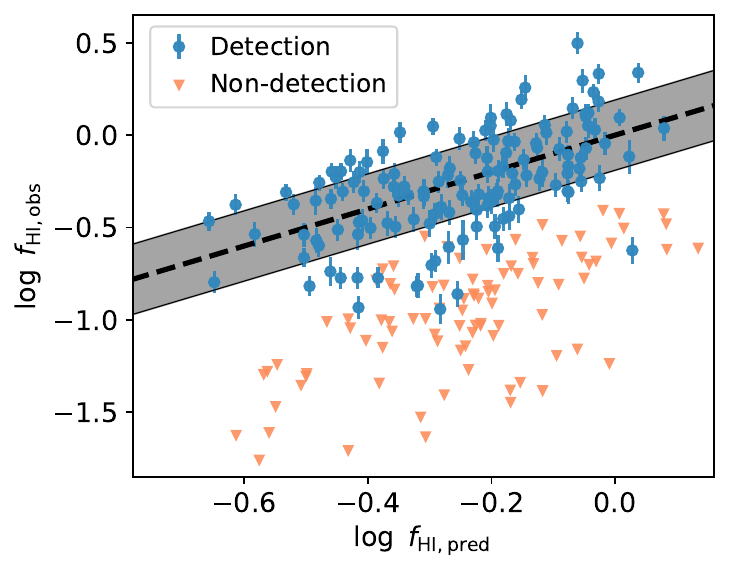}
    \caption{Predicted versus measured H\,{\sc i} gas fraction in control galaxies. The blue points represent galaxies detected in H\,{\sc i}, while the orange triangles indicate non-detections. The dashed line shows the one-to-one relation, and the shaded region denotes the $1\sigma$ intrinsic scatter. The censored data points (orange points) only serve a constraining and regularizing role. As a result, they fall below the model.}
    \label{fig:fh1_mod}
\end{figure}

To predict the H\,{\sc i} gas fraction using the basic properties of galaxies, we modeled it using a linear mixture model of the stellar mass and color based on our control sample. The relation is:
\begin{equation}
    \log f_{\mathrm{HI}}=m_1\log(M_*/M_\odot)+m_2\ (g-r)+b+\sigma_{\mathrm{HI}}\ ,
\label{equ:define_model}
\end{equation}
where $b$ is a constant. Following \citet{Sorgho25}, we considered the normal distribution for detections and the censored distribution for non-detections. The distribution is defined as:
\begin{equation}
f_{Y_\mathrm{cen}}(y) = 
\begin{cases}
\mathcal{N}(\mu, \sigma) & y=M_{\mathrm{HI,\ det}}, \\
1-F_Y(\mu_\mathrm{c}, M_{\mathrm{HI,\ up}}) & y\leq M_{\mathrm{HI,\ up}},
\end{cases}
\end{equation}
where $\mu$ is the model defined by Equation \ref{equ:define_model}. $F_Y(\mu_c, M_{\mathrm{HI,\ up}})$ is the probability that a normally distributed variable with the censored mean $\mu_\mathrm{c}$ is below the H\,{\sc i} mass upper limit $M_{\mathrm{HI,\ up}}$. We use the Markov chain Monte Carlo (MCMC) implementation to fit the relation via P{\scriptsize Y}MC \citep{Abril23}. Figure \ref{fig:mcmc_distri} shows the distributions of posteriors for H\,{\sc i} gas fraction prediction. Each diagonal panel is the marginalized $1$D posterior distribution of the parameters. The $1\sigma$ constrains are $m_1=-0.44\pm0.10$, $m_2=-1.14\pm0.36$, $b=4.67\pm0.80$ and $\sigma_{\mathrm{HI}}=0.19\pm0.01$. A correlation between $m_1$ and $b$ suggests that the model may explain part of the observed trend of $f_{\mathrm{HI}}$ with $M_*$ by invoking a Malmquist bias \citep{Zu20}. The posterior probability of $m_1$ tends to be zero around $m_1=-0.23$, indicating that a negative intrinsic correlation between $f_{\mathrm{HI}}$ and $M_*$ at fixed $g$-$r$ remains useful for interpreting the data. The fitted relation is:
\begin{equation}
    \log f_{\mathrm{HI}}=-0.44\log(M_*/M_\odot)-1.14\ (g-r)+4.67\ .
\label{eq:fh1_model}
\end{equation}
We plot observed and predicted H\,{\sc i} gas fraction in Figure \ref{fig:fh1_mod}. Most of the H\,{\sc i}-detected galaxies (blue points) are located along the one-to-one relation (dashed line). The censored data points (orange points) only serve a constraining and regularizing role and they fall below the model. The Pearson coefficient value is $0.6$ with $p$-value $=0.0$ for H\,{\sc i}-detected galaxies. It indicates that our model can predict H\,{\sc i} gas fraction well.  \\

\begin{deluxetable}{p{1.5cm}cccc}
\setlength{\tabcolsep}{0.5mm}
\label{tab:delta_fh1_median}
\tablewidth{0pt} 
\tablecaption{Median Value of $\Delta f_{\mathrm{HI}}$}
\tablehead{
\colhead{} & \multicolumn{2}{c}{\centering Full Sample} & \multicolumn{2}{c}{\centering H\,{\sc i}-Detections}\\
\cline{2-5}
\colhead{} & \colhead{Median} & \colhead{$95\%$ C.I.} & \colhead{Median} & \colhead{$95\%$ C.I.}
}
\startdata 
Group Galaxies & -0.26 & [-1.15, 0.37] & -0.05 & [-0.69, 0.41] \\
\hline
Control Galaxies & -0.21 & [-1.33, 0.34] & 0.002  & [-0.51, 0.35] \\
\hline
Difference & -0.04 & [-0.18, 0.16] & -0.04 & [-0.14, 0.04] \\
\enddata
\end{deluxetable}

To investigate the difference between the observed data and those predicted by Equation \ref{eq:fh1_model}, we further calculated the offset of H\,{\sc i} gas fraction ($\Delta f_{\mathrm{HI}}$) using: 
\begin{equation}
    \Delta f_{\mathrm{HI}}=\log f_{\mathrm{HI,obs}}-\log f_{\mathrm{HI,pred}}\ .
\label{equ:def_h1}
\end{equation}
We use the Kaplan-Meier (KM) estimator to compute the median $\Delta f_{\mathrm{HI}}$ for the full sample \citep{Feigelson85,Xu98}. And the results are shown in Table \ref{tab:delta_fh1_median}. Figure \ref{fig:h1_def} presents the distribution of $\Delta f_{\mathrm{HI}}$ and the control galaxies are shown using red columns. The median value of $\Delta f_{\mathrm{HI}}$ is $0.002$ dex for H\,{\sc i}-detected control galaxies. For H\,{\sc i} non-detections, the median value is $-0.21$ dex since the upper limits fall below the model prediction. The average standard deviation of the control galaxies is $\sigma_{\mathrm{det}}=0.43$ for H\,{\sc i} detections and $\sigma_{\mathrm{full}}=0.22$ for the full sample. \\

\section{Results}

\subsection{H\textsc{i} gas fraction in group galaxies}

\begin{figure}[h]
    \centering
    \includegraphics[width=0.38\textwidth]{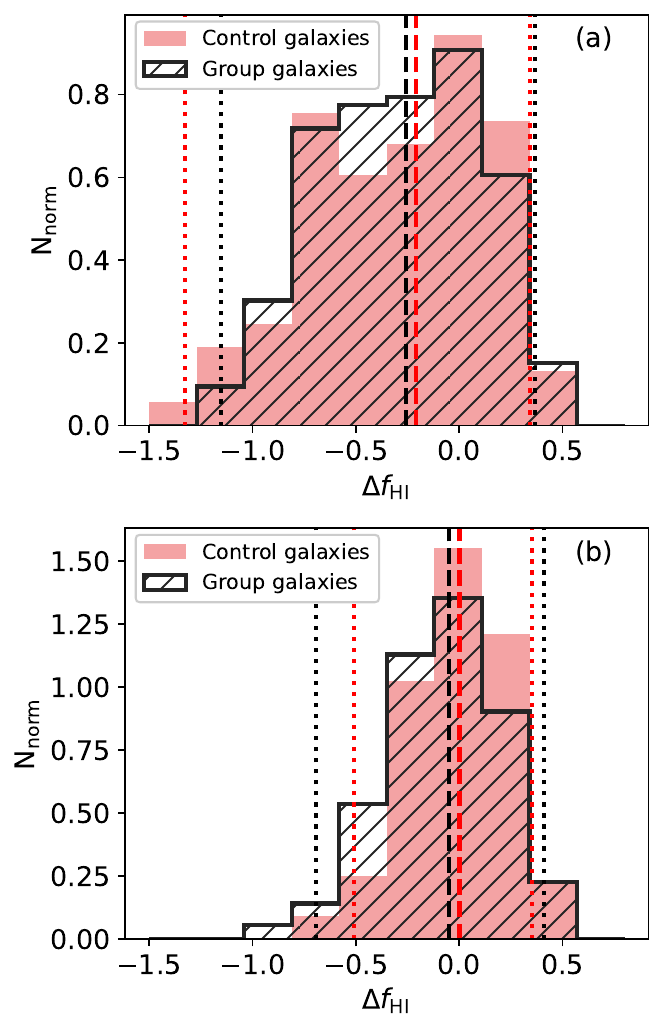}
    \caption{Distribution of $\Delta f_{\mathrm{HI}}$ in full sample (a) and H\,{\sc i}-detections (b). The red and black columns indicate the control galaxies and the group galaxies, respectively. The red and black dashed lines in each panel indicate the median values of the control and group galaxies, respectively. The region between two dotted lines in each panel is $95\%$ confidence interval, shown in red for the control galaxies and black for the group galaxies.}
    \label{fig:h1_def}
\end{figure}

\begin{figure}[h]
    \centering
    \includegraphics[width=0.44\textwidth]{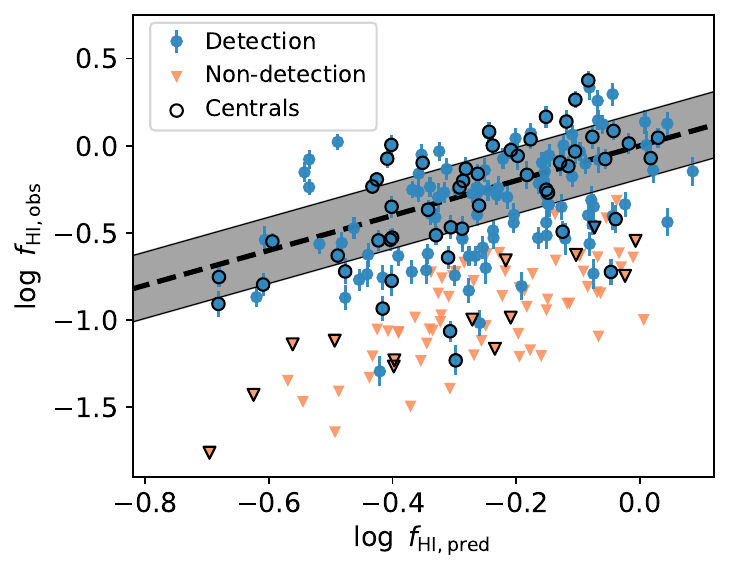}
    \caption{Predicted versus measured H\,{\sc i} gas fraction in group galaxies. The color scheme is the same as Figure \ref{fig:fh1_mod}. Central galaxies are marked by black edges.}
    \label{fig:fh1_group}
\end{figure}

To compare H\,{\sc i} gas fractions in groups with the control, we predicted the H\,{\sc i} gas fractions in group galaxies using Equation \ref{eq:fh1_model}. We show the observed H\,{\sc i} gas fraction versus predicted H\,{\sc i} gas fraction in Figure \ref{fig:fh1_group}. The central galaxies are marked by black edges. We calculated $\Delta f_{\mathrm{HI}}$ using Equation \ref{equ:def_h1}, and the distribution is shown by the black columns in Figure \ref{fig:h1_def}. Using the KM estimator, the median $\Delta f_{\mathrm{HI}}$ is $-0.26$ dex for the full sample and $-0.05$ dex for detections (Table \ref{tab:delta_fh1_median}).  \\

Galaxies with $\Delta f_{\mathrm{HI}}<0$ have H\,{\sc i} content lower than that predicted by the model. Because the control sample includes non-detections, its median H\,{\sc i} fraction can also appear with $\Delta f_{\mathrm{HI}} < 0$. To enable a fair comparison, we therefore use the difference in $\Delta f_{\mathrm{HI}}$ between group and control galaxies to quantify H\,{\sc i} deficiency. Group galaxies with $\Delta f_{\mathrm{HI}}$ values lower than those of the control sample are considered H\,{\sc i} deficient. Figure \ref{fig:h1_def} represents the distribution of $\Delta f_{\mathrm{HI}}$ in full sample (a) and H\,{\sc i} detections (b). Among the group galaxies, $91$ H\,{\sc i}-detected and $126$ in the full sample show $\Delta f_{\mathrm{HI}}$ values lower than the control galaxies. We employ a bootstrap resampling procedure to estimate the difference and $95\%$ confidence interval between the median value of group galaxies and control ones. The difference between group and control galaxies is $-0.04$ dex ($\sim0.5\sigma$) in both cases. Both the H\,{\sc i} detected and the full sample are H\,{\sc i} deficient by $\sim8.8\%$ compared to the control sample. \\

\begin{figure*}[ht]
    \centering
    \includegraphics[width=1\textwidth]{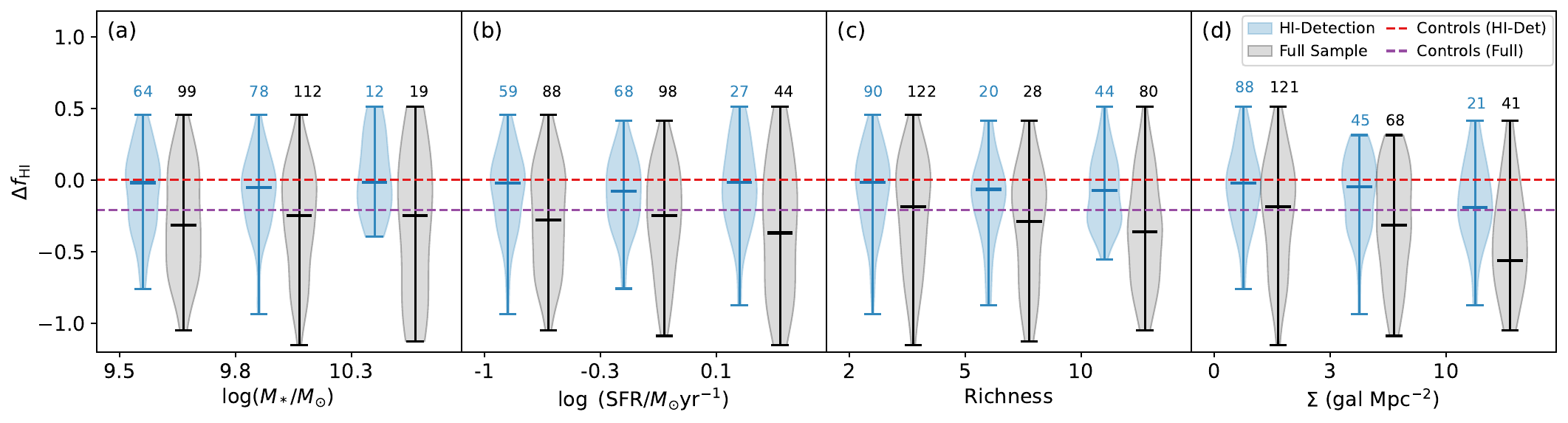}
    \caption{$\Delta f_{\mathrm{HI}}$ of group galaxies versus properties in different bins: stellar mass (a), SFR (b), group richness (c) and local density (d). Black and blue violins are the results of the full sample and H\,{\sc i}-detections, respectively. The small horizontal line in each violin region indicates the median value. The number of galaxies is indicated above the corresponding violin. The median $\Delta f_{\mathrm{HI}}$ of the control sample is shown as a red dashed line for H\,{\sc i}-detections and as a purple dashed line for the full sample, respectively.}
    \label{fig:delta_fh1_binned}
\end{figure*}

\subsection{$\Delta f_{\mathrm{HI}}$ with other properties}

Figure \ref{fig:h1_def} and Table \ref{tab:delta_fh1_median} indicate that group galaxies are H\,{\sc i} modestly deficient relative to the control galaxies. To understand how H\,{\sc i} deficiency in group galaxies correlates with other properties, we plot binned $\Delta f_{\mathrm{HI}}$ against the stellar mass, SFR, group richness and local density in Figure \ref{fig:delta_fh1_binned}(a), (b), (c), and (d), respectively. Specifically, we divided the stellar mass into bins of $[9.5,\ 9.8]$, $[9.8,\ 10.3]$, and $[10.3,\ 10.7]$. These intervals roughly separate low-mass disk galaxies, intermediate-mass disk galaxies, and the massive end of our sample. The SFR was divided into bins of $[-1,\ -0.3]$, $[-0.3,\ 0.1]$, and $[0.1,\ 0.94]$, from almost quenched systems to star-forming galaxies. The group richness was binned into $[2,\ 5]$, $[5,\ 10]$, and $[10,\ 312]$, indicating small groups, developed groups and clusters. The local density was divided into $[0,\ 3]$, $[3,\ 10]$, and $[10,\ 65]$, from loose to compact groups. Figure \ref{fig:delta_fh1_binned} shows the dependence of atomic gas content on both galaxy-intrinsic and environmental parameters. In each violin plot, the small horizontal line marks the median value estimated using KM method. The median $\Delta f_{\mathrm{HI}}$ values of the control sample are shown as dashed lines, with red indicating H\,{\sc i}-detected galaxies and purple representing the full sample, respectively. \\

For the full sample and H\,{\sc i} detections, the median $\Delta f_{\mathrm{HI}}$ is consistently lower than the control sample regardless of stellar mass or SFR. H\,{\sc i} deficiency is more pronounced for the full sample in dense environments (richness $>10$ and $\Sigma>10\ \mathrm{gal\ Mpc^{-2}}$). For the full sample, there is a decrease in deficiency with increasing stellar mass, as suggested by the medians of the first, second, and third violins. \\

\section{Discussion}

\subsection{Central and Satellite Galaxies}

As shown in Figure \ref{fig:fh1_group}, most satellite galaxies lie below the $1\sigma$ error region of the model unlike centrals. We calculated $\Delta f_{\mathrm{HI}}$ for central and satellite galaxies separately (Table \ref{tab:deltafh1_median_cen_sat}). For the full sample, centrals have a median $\Delta f_{\mathrm{HI}}=-0.12$ dex, while satellites have a median of $-0.36$ dex. Compared to the control sample, the differences in $\Delta f_{\mathrm{HI}}$ are $0.13$ dex ($1.4\sigma$) and $0.01$ dex ($0.2\sigma$) for the total and H\,{\sc i}-detected centrals, respectively, corresponding to slight enhancements in H\,{\sc i} gas fraction of $\sim35\%$ and $\sim2.3\%$. By contrast, satellites show differences of $-0.12$ dex ($1.3\sigma$, full sample) and $-0.06$ dex ($1\sigma$, H\,{\sc i}-detected), indicating H\,{\sc i} deficiencies of $\sim24\%$ and $\sim13\%$, respectively. Satellite galaxies tend to have a slight H\,{\sc i} deficiency compared to the central galaxies. \\

\begin{figure*}[ht]
    \centering
    \includegraphics[width=0.85\textwidth]{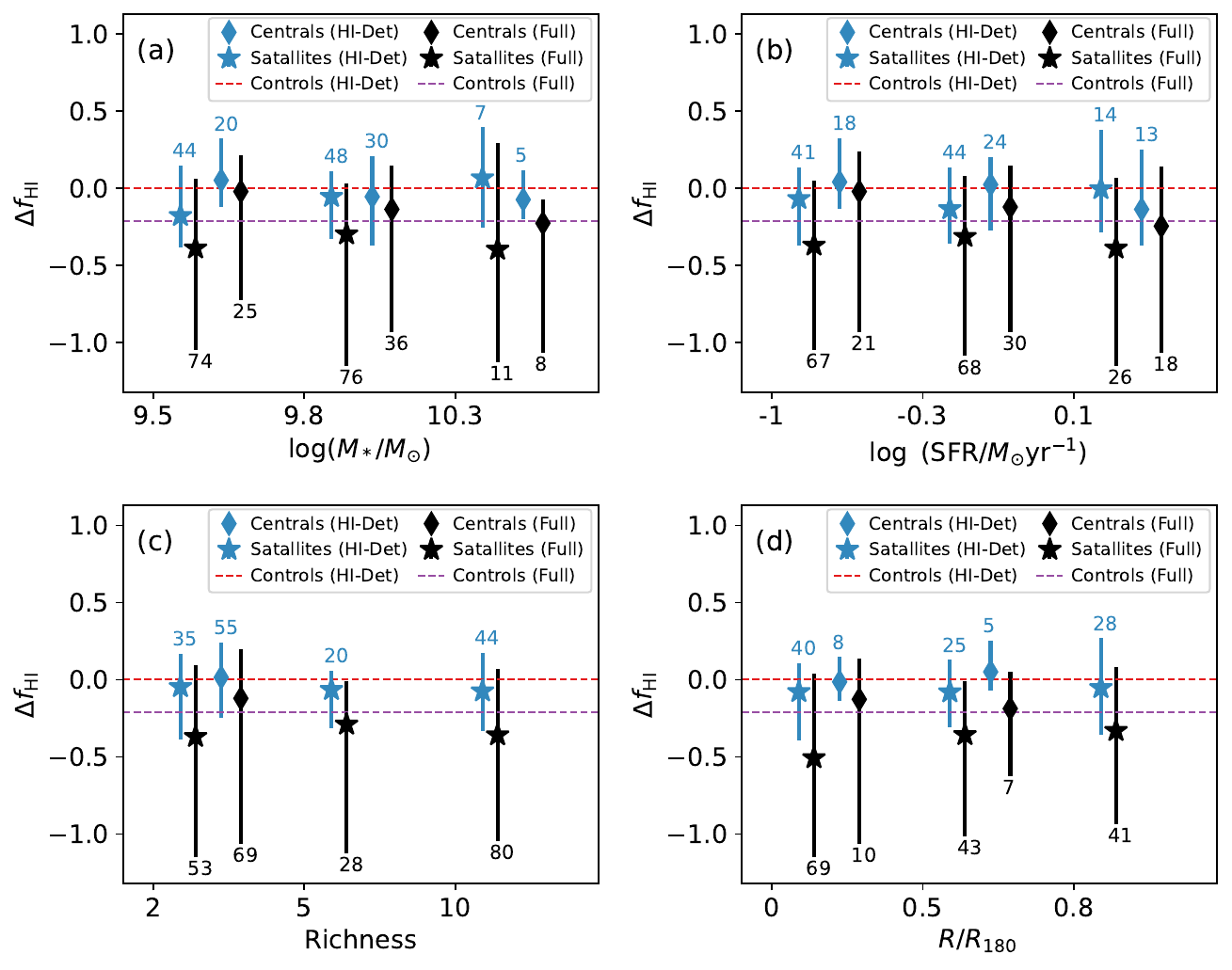}
    \caption{Median $\Delta f_{\mathrm{HI}}$ of central and satellite galaxies against stellar mass (a), SFR (b), richness (c), and $R/R_{180}$ (d). Black and blue color indicates the results of the full sample and H\,{\sc i}-detections, respectively. The error bars are $16\%\sim84\%$ confidence interval. Central and satellite galaxies are shown using diamonds and stars, respectively. The median $\Delta f_{\mathrm{HI}}$ of the control sample is shown as a red dashed line for H\,{\sc i} detections and as a purple dashed line for the full sample, respectively. The number of related galaxies is shown in each point.}
    \label{fig:delta_fh1_cen_sat}
\end{figure*}

We plot the binned results of median $\Delta f_{\mathrm{HI}}$ of central and satellite galaxies in Figure \ref{fig:delta_fh1_cen_sat}. The central galaxies in our sample have richness below $5$, primarily because our selection criterion removes close companions within $2'.9$. For satellite galaxies, our selection criteria may exclude those located close to the centrals. However, the satellites at large projected distances from the central can still be included, even in richer groups. As shown in Figure \ref{fig:delta_fh1_cen_sat}(c), these satellites can reside in groups with richness $>5$. The error bars are $16$-$84\%$ confidence interval calculated using the KM estimator. The central and satellite galaxies are shown using diamonds and stars, respectively. The black and blue dashed line indicates the median $\Delta f_{\mathrm{HI}}$ of the total control sample and that of the H\,{\sc i} detected control sample. In addition to the stellar mass, SFR, and group richness, we also consider the projected distance between each satellite galaxy and the central galaxy of its host halo. For centrals, we use the distance to the nearest satellite, normalized by the group halo radius $R_{180}$. We calculate $R_{180}$ via \citet{Yang07}:
\begin{equation}
    R_{180}=1.26\ h^{-1}\ \mathrm{Mpc}(\frac{M_h}{10^{14}h^{-1}M_\odot})^{1/3}(1+z_{\mathrm{group}})^{-1}\ ,
\end{equation}
where $M_h$ is the halo mass adopted directly from the catalog of \citet{Yang07}. But the group halo mass is only available for massive halos with $M_h\ge10^{11.5}M_{\odot}$. Consequently, only $171$ galaxies in our sample have available halo mass estimates for their host halos. We plot binned $\Delta f_{\mathrm{HI}}$ against $R/R_{180}$ for central and satellite galaxies in Figure \ref{fig:delta_fh1_cen_sat}(d). \\

In H\,{\sc i}-detected galaxies, deficiency is more evident in low mass ($M_*<10^{9.8}M_\odot$) satellites (difference $\sim1\sigma$). For the full sample and H\,{\sc i} detections, the deficiency is not obvious with the SFR (difference $<1\sigma$). Among satellites, no clear dependence of H\,{\sc i} deficiency on group richness is shown. H\,{\sc i} deficiency becomes more pronounced as satellites approach the group center ($R/R_{180}<0.5$) for the full sample (difference $\sim1\sigma$). By contrast, H\,{\sc i}-detected central galaxies exhibit no significant deficiency in H\,{\sc i} fraction regardless of the location of their nearest satellite. \\

\begin{deluxetable}{p{1.7cm}cccc}
\setlength{\tabcolsep}{0.5mm}
\label{tab:deltafh1_median_cen_sat}
\tablewidth{0pt}
\tablecaption{\\Median Value of $\Delta f_{\mathrm{HI}}$ in Centrals and Satellites}
\tablehead{
\colhead{} & \multicolumn{2}{c}{\centering Full Sample} & \multicolumn{2}{c}{\centering H\,{\sc i}-Detections}\\
\cline{2-5}
\colhead{} & \colhead{Median} & \colhead{$95\%$ C.I.} & \colhead{Median} & \colhead{$95\%$ C.I.}
}
\startdata 
Centrals & -0.12 & [-1.06, 0.37] & 0.02 & [-0.73, 0.39] \\
\hline
Difference to Controls & 0.13 & [-0.02, 0.33] & 0.01  & [-0.12, 0.10] \\
\hline
Satellites & -0.36 & [-1.15, 0.33] & -0.07  & [-0.64, 0.40] \\
\hline
Difference to Controls & -0.12 & [-0.31, 0.08] & -0.06  & [-0.22, 0.02] \\
\enddata
\end{deluxetable}
\subsection{Comparison with previous studies}

Several studies have investigated the H\,{\sc i} deficiency in HCGs. \citet{Jones23} analyzed VLA H\,{\sc i} observations of $38$ HCGs ($3$ non-detections), and \citet{Sorgho25} observed $6$ HCGs ($3$ non-detections) with MeerKAT. Both studies found that HCGs generally contain less H\,{\sc i} than predicted by scaling relations, with the deficiency being most pronounced in systems where the atomic gas is entirely removed (H\,{\sc i} detected in the tidal tail) or undetected. HCGs have low richness, typically containing $4$-$10$ members, but they are compact with local density above $100\ \mathrm{gal\ Mpc^{-2}}$ \citep{Hickson82}. Although our sample does not contain any HCG, it is still important to investigate H\,{\sc i} content for galaxies residing in higher local density environments. Therefore, we plot $\Delta f_{\mathrm{HI}}$ versus local density in Figure \ref{fig:delta_fh1_binned}(d). H\,{\sc i} deficiency becomes significant at $\Sigma>10\ \mathrm{gal\ Mpc^{-2}}$, where the H\,{\sc i}-detected group sample has a median $\Delta f_{\mathrm{HI}}=-0.19$ dex, and the full sample exhibits a stronger deficiency with a median $\Delta f_{\mathrm{HI}}=-0.56$ dex. Compared to the isolated galaxies, $\Delta f_{\mathrm{HI}}$ is lower by $\sim0.5\sigma$ in both sample. As galaxies approach the group center, the local density increases. In such dense regions, ram pressure stripping by the hot intragroup medium or circumgalactic medium can remove gas, leading to H\,{\sc i} deficiency \citep{Rasmussen06,Rasmussen12,Moon19,Kolcu22}. Meanwhile, for group galaxies with richness below $5$ (particularly for galaxy pairs), tidal interaction with close companions is still efficient to remove gas \citep{Huang24}. Figure \ref{fig:delta_fh1_cen_sat}(d) also shows that satellite galaxies located closer to their central galaxies exhibit higher H\,{\sc i} deficiency. In addition, satellite galaxies with low stellar mass ($M_* < 10^{9.8}M_{\odot}$) remain modest H\,{\sc i}-deficient ($0.5\sigma$, Figure \ref{fig:delta_fh1_cen_sat}(a)). Low-mass systems have a shallow potential well. In low-mass systems or regions with high local density, environmental processes such as ram pressure stripping and tidal interactions are likely to remove gas efficiently, leading to reduced atomic gas content. \\

\citet{Janowiecki17} studied H\,{\sc i} content of $161$ central galaxies using Arecibo data. Approximately $80\%$ of their groups have group members less than $4$. They found that they generally have higher H\,{\sc i} content than isolated galaxies. They also reported that centrals with more distant nearest satellites show stronger H\,{\sc i} enhancement. In our work, Figure \ref{fig:delta_fh1_cen_sat}(a) shows a weak enhancement ($>0.5\sigma$) in low-mass central galaxies ($M_*<10^{9.8}M_\odot$) for the full sample. The cold gas accreted from the cosmic web or recent H\,{\sc i}-rich minor mergers into the low-mass groups can not get heated, unlike in larger groups. As a result, the cold gas may survive for a long period in these low-mass groups. Further detailed investigation of low-mass groups is required to confirm and characterize the tentative enhancement in gas content.  \\

\citet{Brown23} studied star formation in $33$ Virgo cluster satellite galaxies and found that H\,{\sc i}-poor galaxies have a reduced SFR surface density compared to H\,{\sc i}-normal cluster or field galaxies. Ram pressure stripping plays a key role in depleting atomic gas reservoirs and quenching star formation. The reduced atomic gas reservoirs may link to lower SFR in these galaxies.  \\

As shown in Figure \ref{fig:delta_fh1_cen_sat}(a) and (b), central galaxies display trends that differ from those of satellites. \citet{Janowiecki17} reported $\mathrm{H_2}$ observations for $7$ central galaxies with stellar masses below $10^{10}M_{\odot}$, $6$ of which display higher $\mathrm{H_2}$ fractions than isolated galaxies of similar mass. Their findings suggest that low-mass central galaxies may simultaneously exhibit elevated H\,{\sc i} fractions, enhanced $\mathrm{H_2}$ content, and increased SFR. In this work, the average star formation rate is  $\log(\mathrm{SFR/M_\odot\mathrm{yr^{-1}}})=-0.11$ for centrals and $\log(\mathrm{SFR/M_\odot\mathrm{yr^{-1}}})=-0.23$ for satellites. In contrast to satellites, which tend to lose H\,{\sc i} through stripping, central galaxies may accrete cold gas from their surroundings as mentioned before, sustaining or even boosting their star formation activity. \\ 

\section{Summary}
In this work, we investigated the H\,{\sc i} content of group galaxies using FASHI data and the SDSS DR$7$ catalog. Our sample contains $230$ group galaxies from \citet{Yang07}, including $154$ H\,{\sc i} detections. Of these, $69$ are central galaxies ($55$ detections) and $161$ are satellites ($99$ detections). For comparison, we constructed a control sample of $230$ isolated galaxies ($140$ detections), matched in stellar mass and color to the group galaxies. We use the control sample to establish a scaling relation between H\,{\sc i} gas fraction and optical properties. We then applied the relation to predict the H\,{\sc i} content of the group galaxies. We calculated $\Delta f_{\mathrm{HI}}$ using Equation \ref{equ:def_h1}. Galaxies with $\Delta f_{\mathrm{HI}}$ lower than those of the control galaxies are considered H\,{\sc i} deficient. The main conclusions of this study are summarized below: \\

1. We estimate the median $\Delta f_{\mathrm{HI}}$ using the KM estimator. For the control galaxies, the median is $-0.21$ dex ($95\%$ CI [$-1.33,\ 0.34$]) with $\sigma=0.43$ for the full sample. It is $0.002$ dex ($95\%$ CI [$-0.51,\ 0.35$]) with $\sigma=0.22$ for the H\,{\sc i}-detected sample. For the full group galaxies, the median is $-0.26$ dex ($95\%$ CI [$-1.15,\ 0.37$]) with $\sigma=0.39$, while it is $-0.05$ dex ($95\%$ CI [$-0.69,\ 0.41$]) with $\sigma=0.28$ for H\,{\sc i}-detected group galaxies. Both the entire group galaxies and H\,{\sc i}-detected group galaxies show mild H\,{\sc i} deficiency compared to the control sample. The difference of $\Delta f_{\mathrm{HI}}$ between group and control galaxies is $-0.04$, corresponding to an $\sim8.8\%$ lower H\,{\sc i} content in group galaxies compared to isolated systems. \\

2. $\Delta f_{\mathrm{HI}}$ is independent of the stellar mass and SFR. Low $\Delta f_{\mathrm{HI}}$ is more pronounced but still modest ($\sim0.5\sigma$) in the full sample in denser environments (richness $>10$ or $\Sigma>10\ \mathrm{gal\ Mpc^{-2}}$).  \\

3. Compared to the control sample, the differences in $\Delta f_{\mathrm{HI}}$ are $0.13$ dex ($1.4\sigma$) and $0.01$ dex ($0.2\sigma$) for the total and H\,{\sc i}-detected centrals, respectively. By contrast, satellites show differences of $-0.12$ dex ($1.3\sigma$, full sample) and $-0.06$ dex ($1\sigma$, H\,{\sc i}-detected). Centrals likely accrete gas from minor mergers or the cosmic web, enhancing their H\,{\sc i} reservoirs, consistent with the findings of \citet{Janowiecki17}. In contrast, satellite galaxies tend to lose gas when entering denser regions (e.g., ram pressure stripping, tidal interactions), resulting in H\,{\sc i} depletion.  \\

In conclusion, we have investigated H\,{\sc i} gas content of group galaxies using recent FASHI data, providing a systematic view of atomic gas evolution in galaxy groups. Group galaxies show a mild deficit in H\,{\sc i} relative to control systems, with the deficiency becoming more significant in denser regions (richness $>10$ or $\Sigma > 10\ \mathrm{gal\ Mpc^{-2}}$). The satellite galaxies can lose gas due to the ram pressure stripping of the hot intragroup medium or tidal interactions with companions. Low-mass central galaxies may be accreting gas from their surroundings. Our tentative results still need further investigation. Future observations focusing on molecular gas in group galaxies, based on our studies, will provide deeper insight into the regulation of cold gas and star formation in group environments. In addition, future X-ray observations will help characterize the hot intragroup medium, enabling a more precise assessment of halo stripping processes. \\

This work is supported by the National Natural Science Foundation of China under Nos. 11890692, 12133008, 12221003, 12550002. We acknowledge the science research grants from the China Manned Space Project with No. CMS-CSST-2021-A04 and No. CMS-CSST-2025-A10. QY was supported by the European Research Council (ERC) under grant agreement No. 101040751. This work uses the data from FASHI project. FASHI made use of the data from FAST (Five-hundred-meter Aperture Spherical radio Telescope).  FAST is a Chinese national mega-science facility, operated by National Astronomical Observatories, Chinese Academy of Sciences. \\

Funding for the Sloan Digital Sky Survey IV has been provided by the Alfred P. Sloan Foundation, the U.S. Department of Energy Office of Science, and the Participating Institutions. SDSS-IV acknowledges support and resources from the Center for High Performance Computing  at the University of Utah. The SDSS website is \url{www.sdss.org}. \\

SDSS-IV is managed by the Astrophysical Research Consortium for the Participating Institutions of the SDSS Collaboration, including the Brazilian Participation Group, the Carnegie Institution for Science, Carnegie Mellon University, Harvard-Smithsonian Center for Astrophysics, the Chilean Participation Group, the French Participation Group, Instituto de Astrof\'isica de Canarias, The Johns Hopkins University, Kavli Institute for the Physics and Mathematics of the Universe (IPMU) / University of Tokyo, the Korean Participation Group, Lawrence Berkeley National Laboratory, Leibniz Institut f\"ur Astrophysik Potsdam (AIP),  Max-Planck-Institut f\"ur Astronomie (MPIA Heidelberg), Max-Planck-Institut f\"ur Astrophysik (MPA Garching), Max-Planck-Institut f\"ur Extraterrestrische Physik (MPE), National Astronomical Observatories of China, New Mexico State University, New York University, University of Notre Dame, Observat\'ario Nacional / MCTI, The Ohio State University, Pennsylvania State University, Shanghai Astronomical Observatory, United Kingdom Participation Group, Universidad Nacional Aut\'onoma de M\'exico, University of Arizona, University of Colorado Boulder, University of Oxford, University of Portsmouth, University of Utah, University of Virginia, University of Washington, University of Wisconsin, Vanderbilt University, and Yale University. \\

$Software:$ Astropy \citep{astropy13}, PyMC \citep{Abril23}, SciPy \citep{scipy}, TOPCAT \citep{TOPCAT}.

\bibliography{references}{}
\bibliographystyle{aasjournal}

\end{CJK*}
\end{document}